\documentclass[11pt,a4paper]{article}
\usepackage{jcappub}

\title{Precision cosmology and $^7 Li$ data}

\author[a,b]{G. La Vacca,} 
\author[a]{A. Valotti}
\affiliation[a]{Physics Department G.~Occhialini, Milano-Bicocca University\\
Piazza della Scienza 3, 20126 Milano, Italy}
\affiliation[b]{I.N.F.N., Sezione di Milano-Bicocca,\\ 
Piazza della Scienza 3, 20126 Milano, Italy}

\author[c,d]{and S. A. Bonometto} 
\affiliation[c]{Physics Department, Astrophysics Section, Trieste University,\\ 
Via Tiepolo 12, 34131 Trieste, Italy}
\affiliation[d]{I.N.F.N., Sezione di Trieste,\\
Via Valerio 2, 34127 Trieste, Italy}

\emailAdd{lavacca@mib.infn.it}
\emailAdd{a.valotti@campus.unimib.it}
\emailAdd{bonometto@oats.inaf.it}

\abstract{At variance from $^2H$, $^3 He$ and $^4 He$ abundances, $^7
Li$ abundance data yield an extra constraint to cosmological
parameters, in top of those deriving from CMB, BAO, SNIa or $H_0$
data. This constraint, often disregarded, would favor smaller
$\Omega_b h^2$ values, also indicating a preference for Dark Energy
state equations well in the phantom regime, simultaneously softening
the upper limit on the sum of neutrino masses, up to $\sim 1.6~$eV.}

\keywords{Dark energy theory, dark matter, cosmological neutrinos,
neutrino properties, cosmology of theories beyond the SM}
\arxivnumber{...}

\begin{document}
\maketitle

\section{Introduction}
Big-Bang Nucleosynthesis (BBN) is a pillar of evolutionary
cosmology. It is however known that ``precision'' cosmology is not in
complete agreement with BBN predictions on $\Omega_b h^2$ ($\Omega_b:$
baryon density parameter, $h:$ Hubble parameter in units of 100
(km/s)/Mpc$\, \, $). Agreement is however recovered if $^7 Li$
abundance data are disregarded.

$^7 Li$ is the heaviest and most scanty nuclide considered in BBN, but
the reaction network leading to is hardly questionable, while its
abundance data, resumed in the next Section, are sound. It seems
therefore significant to debate what changes in cosmological parameter
estimates if $^7 Li$ abundance data are taken on the same foot as
other nuclides.

BBN aims at predicting the abundances of $^2H$, $^3 He$, $^4 He$, and
$^7 Li$ in terms of a single parameter:
\begin{equation}
\eta_{10} = 10^{10} n_b/n_\gamma = 2.7349 \times 10^2~\Omega_b h^2~.
\label{eta10}
\end{equation}
Here $ n_b$ and $n_\gamma $ are baryon and CMB photons number
densities, respectively. Owing to baryon conservation and assuming no
substantial entropy input between BBN and today, their ratio has not
changed since then. Data on nuclide abundances will be taken here from
recent review papers by Matteucci \cite{b1} and Iocco et
al. \cite{Iocco}, while the network of reactions used for predictions
is accurately illustrated in \cite{Iocco}.

Tests will be based on WMAP7 data \cite{komatsu2}. We shall re-analyse
them by using the {\sc CosmoMC} routines \cite{cosmomc} to gauge model
likelihood. Together with CMB data, Baryon Acoustic Oscillation (BAO)
data \cite{Percival:2009xn}, as well as either $H_0$
\cite{Riess:2009pu} or SNIa data \cite{Amanullah:2010vv}, as suitably
specified, will be considered.

Our analysis will focus on the Dark Energy (DE) state equation,
however assumed to take a scale independent value $w$. Neutrino masses
will be considered through the parameter $M_\nu = \sum_\nu m_\nu$ and
neglecting their mass hierarchy. A
Friedmann-Lema\^itre-Robertson-Walker metric is assumed, with flat
space section.

\begin{figure}
\begin{center}
\includegraphics[height=11.cm,angle=0]{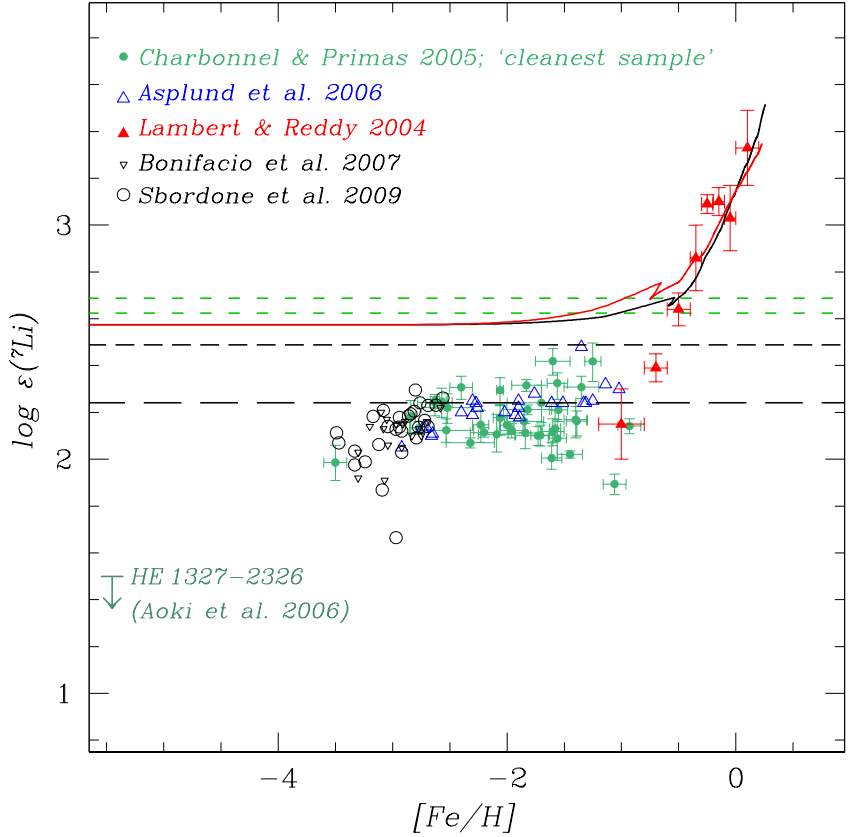}
\end{center}
\caption{Observational $^7Li$ values given by various authors (as in
the frame). The long dashed black horizontal line is the average
$^7Li$ estimate, taken by \cite{Iocco} and used though this paper. The
short dashed black horizontal line shows the abundance allowing a best
fit among $^7Li$ and other nuclide primeval abundances (see text). The
shortest dashed green horizontal lines are the edges of the 1-$\sigma$
abundance interval consistent with WMAP7 (and related) data. The red
and black solid lines show the results of galactic evolution models
assuming a primeval $^7Li$ abundance at 2-3$\sigma$'s below the best
fit of WMAP7 (and related) data with $\Lambda$CDM models.}
\label{matteucci}
\end{figure}

\section{A quick review on \texorpdfstring{$^7Li$}{7Li} stellar abundance}
The history of primordial $^7Li$ abundance determination starts from
the discovery of the Spite \& Spite plateau \cite{spite}. They found
that the $^7Li$ abundance, when the warmest metal-poor dwarfs were
considered, exhibited no appreciable metallicity dependence,
suggesting then a primordial value $\log \epsilon(^7Li) \equiv
[^7Li/H] = 2.05 \pm 0.15$. Here $[X/H] = 12+\log_{10} (X/H) $, $(X/H)$
being the ratio of number densities.

Recent data, using $Fe$ abundances as metallicity indicator, confirm
the plateau for $-3.0 < [Fe/H] < -1.0$ but extend to even lower
metallicities where a mild further $^7Li$ decrease is detected. This
is shown also in Figure \ref{matteucci}, extending the outputs from
\cite{b1}. Observational abundances of $^7Li$, obtained by
\cite{l1,c1, a1,b2,sb}, plus the (controversial) limits in \cite{ao}
are plotted vs.~$[Fe/H]$ (1-$\sigma$ error bars).
\begin{figure}
\begin{center}
\includegraphics[height=11.cm,angle=0]{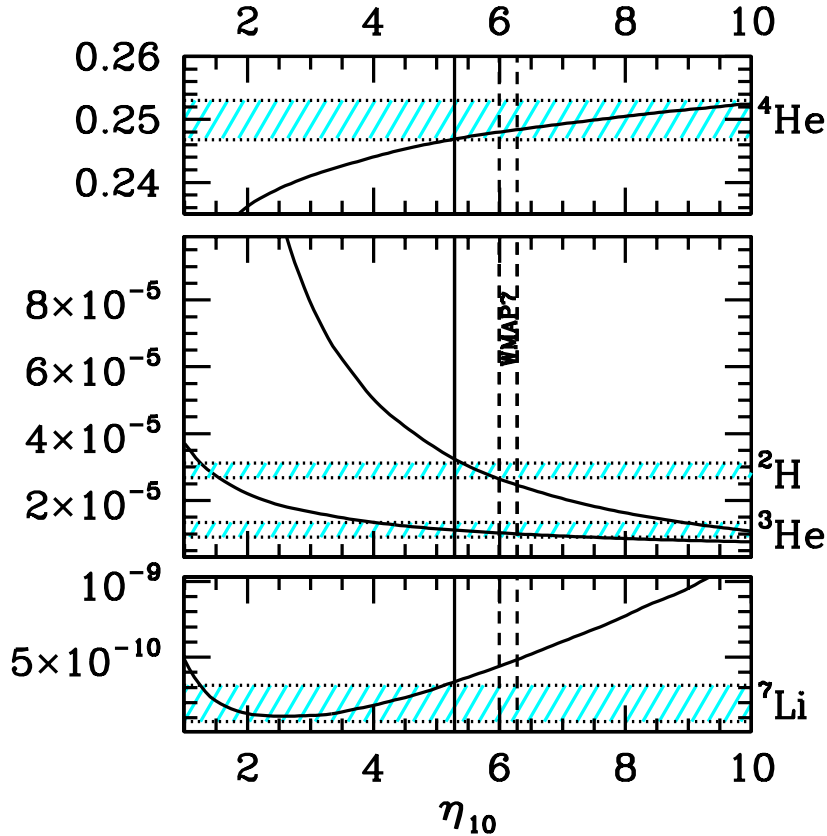}
\end{center}
\caption{Observational abundances of the lightest 4 nuclides produced
in BBN, {\it vs.}~BBN predictions. The horizontal cyan dashed bands
yield 1-$\sigma$ observational limits. 1-$\sigma $ WMAP7 (plus BAO
and $H_0$) data are comprised between the two vertical dashed lines.
The solid vertical line is the best fit with all light nuclide data.
It meets $^3He$ and $^4He$ bands well within 1-$\sigma$, while it
keeps at $\sim 1.3~\sigma$'s from both $^2H$ and $^7Li$ data.}
\label{all1}
\end{figure}
In the same Figure the results of evolutionary models of \cite{rm} are
shown. They assume a primeval $^7Li$ abundance as is predicted for a
value of $\eta_{10}$ lying at $\sim 2-3~\sigma$'s below its best fit
for $\Lambda$CDM models, when using WMAP7 and related data. In top of
it they add the $^7Li$ contributions deriving from low mass stars,
Galactic Cosmic Rays (GCR), novae and Asymptotic Giant Branch (AGB)
stars (ranging around $\sim 41\, \%$, $\sim 25\, \%$, $\sim 9\, \%$
and $\sim 0.5\, \%$, respectively). In this way the high metallicity
rise observed by \cite{l1} is met, so confirming that galactic
evolution allows a fair understanding of the stellar production of
$^7Li$, but hardly meets its values in the Spite \& Spite plateau, let
alone lower metallicity values. In the plot we added the $^7Li$
abundance band consistent with WMAP7 and related datasets (its
1-$\sigma$ edges are the shortest green lines) and a dashed line
representing the average of the eight $^7Li$ estimates quoted in
\cite{Iocco}
\begin{equation}
(^7Li/H) = 1.86 ^{+1.29}_{-1.09}\, \times 10^{-10}~;
\label{obse}
\end{equation}
here the errors are the half-width of the nearly-Gaussian fitting the
above eight estimates; statistical error on each single estimate are
significantly smaller.

\section{BBN predictions on \texorpdfstring{$\eta_{10}$}{eta10} 
and \texorpdfstring{$\Omega_b$}{omb}} 

Letting apart cosmological data analysis, we may look at nuclide data
only and seek the baryon density best approaching a general agreement
(g.a.) among all nuclide abundances.

Within 1-$\sigma$ this is impossible, but, at $\sim 1.3~\sigma$'s from
central values, both for $^7 Li$ and $^2 H$, a g.a. is met (see Figure
\ref{all1}). This however requires $\eta_{10} = 5.369$, i.e. $\Omega_b
h^2 = 1.963 \times 10^{-2}$, a value significantly below the best fit
to WMAP7 and related data.

Moreover, by looking at Figure \ref{matteucci}, we have a visual
feeling of the displacement of such 1.3-$\sigma$ g.a. point (short
dashed black line) from some $^7 Li$ abundance datasets. This point
more or less coincides with observational top estimates of $^7 Li$
abundance and one can hardly escape the impression that, in order to
agree on this point, one must still invoke some unknown --- although
mild --- mechanism for $^7Li$ depletion in old stars.

The formal significance of 1.3~$\sigma$'s is however clear, and tells
us that the heavy vertical line shown in Figure \ref{all1} has a
likelihood of some percents. Hereafter we shall refer to the short
dashed line as ``BBN g.a.~line''. We can also easily determine the
interval where inside we have the 99$\, \%$ of probability to match
all nuclide estimates: $5.25 < \eta_{10} < 5.80$, i.e. $1.92 \times
10^{-2} < \Omega_b h^2 < 2.12 \times 10^{-2}$; hereafter we shall
refer to this interval as ``BBN g.a.~band''.

The asymmetry of the band is due to the different width of the
1-$\sigma$ intervals for $^2H$ and $^7Li$ and to the faster rise of
$^2H$ prediction curve in respect to $^7Li$.  Because of such
asymmetry, $\langle \Omega_b h^2 \rangle = 2.005 \times 10^{-2}$.
\begin{figure}
\begin{center}
\includegraphics[width=11.cm,angle=0]{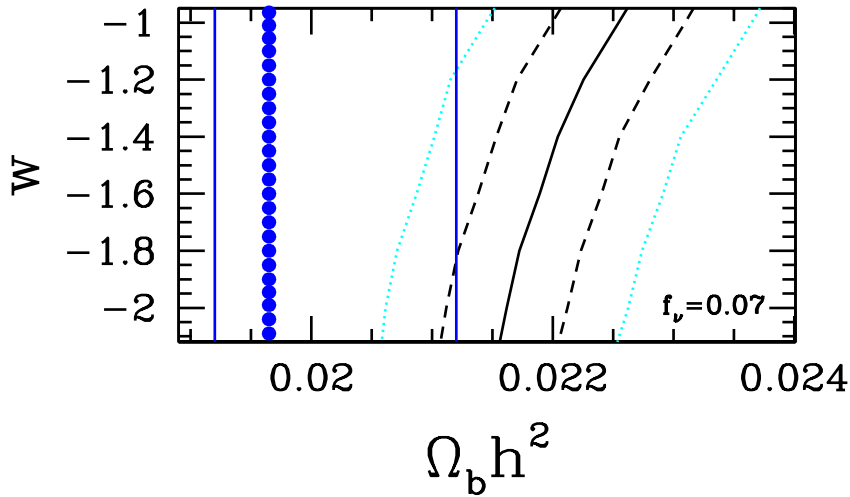}
\end{center}
\caption{65$\, \%$ and 95$\, \%$ likelihood limits on the $\Omega_b
h^2$ parameter (derived from WMAP7, BAO and $H_0$ data and obtained by
fixing $w < -1$ values) are compared with the BBN g.a.~line (vertical
blue dashed line) and band (limited by vertical blue solid lines).
Decreasing $w$ improves the comparison and models with $w < -1.8$
exhibit an overlap between the 1-$\sigma$ likelihood interval and the
BBN g.a.~band, including $^7 Li$.}
\label{datw1}
\end{figure}
In Figure \ref{all1} we also show the 1-$\sigma$ band ($6.031 <
\eta_{10} < 6.326$, i.e. $2.207 < 10^2\, \Omega_b h^2 < 2.313 $),
obtainable by applying {\sc CosmoMC} to WMAP7 data together with BAO
and $H_0$ constraints, assuming a $\Lambda$CDM cosmology; the plot
visually confirms that the ``BBN g.a.~line'' lies at $\sim 6\, \,
\sigma$'s from WMAP and related data best-fit value.

The same plot can be read in a complementary way, by observing that
the theoretical predictions on $^7 Li$ abundance meet the $\eta_{10}$
value best-fitting WMAP7 and related data at $\sim 2.8~ \sigma$'s
(with reference to the data mean and standard deviation shown in
eq.~\ref{obse}).

This is consistent with the absence of overlap between the ``BBN
g.a.~band'' and the $\Lambda$CDM interval; accordingly, the likelihood
for $^7 Li$ data and $\Lambda$CDM agreement is $\ll 0.1 \, \%$. In
turn, this shows that agreement would be re-approached if cosmological
predictions are displaced downwards by $\sim 4\, \%$.

In the era of precision cosmology this is a non-negligible shift.
However, here we show that, if we abandon the assumption that the DE
state parameter $w \equiv -1~,$ a downwards shift of such order is
obtainable.

In the very WMAP Cosmological Parameters
matrix\footnote{http://lambda.gsfc.nasa.gov/product/map/dr4/parameters.cfm}
one can easily check that lower limits on $\Omega_b h^2$ can be
relaxed, if one extra degree of freedom is opened and $w < -1$ is
allowed. We then performed a number of tests, to determine the
best-fit $\Omega_b h^2$ interval with fixed $w < -1$ values, gradually
moving away from $\Lambda$CDM. Fits used {\sc CosmoMC}, constantly
referring to WMAP7, BAO, and $H_0$ data. Their results are shown in
Figure \ref{datw1}.

Fits were also performed with different assumptions on neutrino mass
and we shall further comment on this point in the next section. In
Figure \ref{datw1} neutrinos account for a fraction $f_\nu = 0.07$ of
dark matter mass.

Figure \ref{datw1} shows that decreasing $w$ improves the match
between BBN and WMAP7. Models with $w < -1.8$ exhibit an overlap
between the BBN g.a.~band and the 1-$\sigma$ likelihood interval
obtained by considering WMAP7, BAO's and $H_0$ data.
\begin{figure}
\begin{center}
\includegraphics[height=9.cm,angle=0]{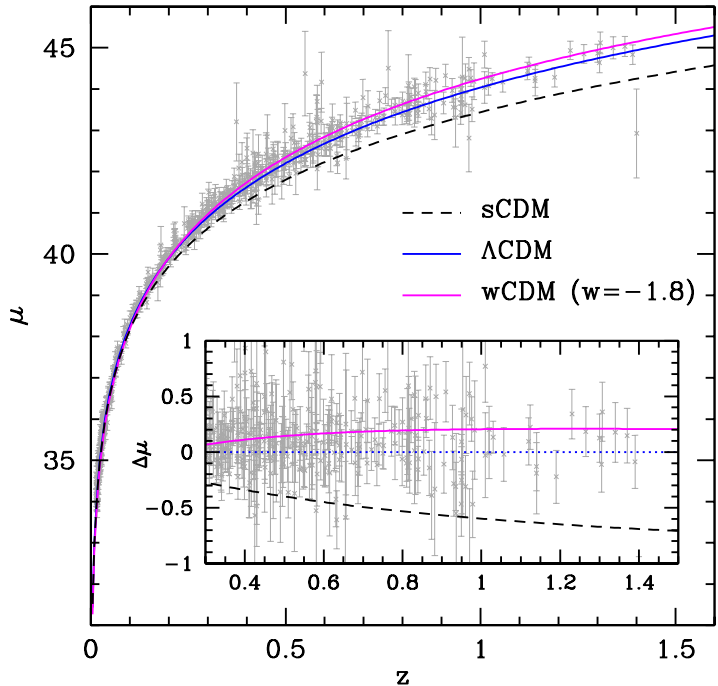}
\end{center}
\caption{Distance moduli of Union2 SNIa data are plotted vs.~different
model predictions.}
\label{HD}
\end{figure}

Let us also outline that results are marginally different if SNIa data
replace $H_0$ data; SNIa and $H_0$ data, however, are not independent
and should not be simultaneously used. In Figure \ref{HD} we however
show the distance modulus $\mu = m-M$, obtained for $\Lambda$CDM and
other cosmologies, against observational distance moduli provided by
the Union2 Supernova Cosmology Project \cite{Amanullah:2010vv}. The
upper solid (magenta) curve, in the main frame, is obtained for $w =
-1.8$. $\Delta \mu$'s are also plotted, in respect to a $\Lambda$CDM
model. There can be scarce doubts that a model with $w = -1.8$ is in
perfect agreement with SNIa data.

\section{Phantom \texorpdfstring{$w$}{w} and neutrinos}
There is a wide literature on models with $w \ll -1$, following the
original proposal by \cite{caldwell} of an anomalous kinetic energy
for the DE scalar field. The main physical context where such {\it
phantom} DE was advocated where the limits on neutrino ($\nu$) masses
(see, e.g., \cite{wsmall}).

\begin{figure}
\begin{center}
\includegraphics[height=10.cm,angle=0]{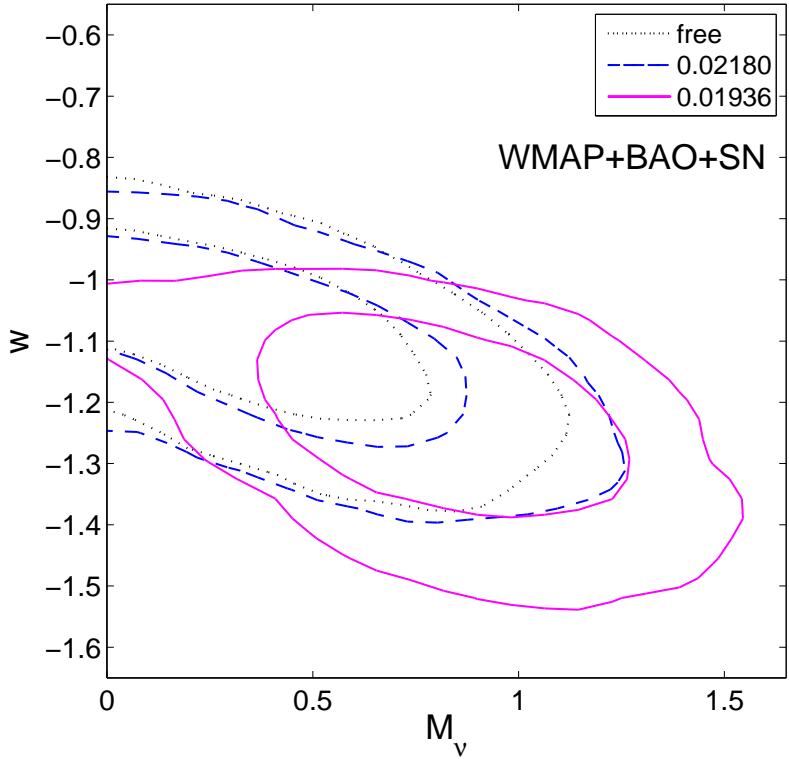}
\end{center}
\caption{1- and 2-$\sigma$ likelihood contours on the $M_\nu$-$w$
plane, when small $\Omega_b h^2$ values are set. The dotted curves
update \cite{komatsu} results, by replacing WMAP5 by WMAP7 data.}
\label{mnu}
\end{figure}

Let then $M_\nu = \sum_\nu m_\nu$ be the sum of the $\nu$-mass
eigenvalues, for $\nu$'s belonging to the usual three particle
families. The point we wish to make here is that, if we force
$\Omega_b h^2$ to shift to values smaller than those obtainable by
freely best fitting cosmological data, so to approach $^7Li$
constraints, we simultaneously obtain two results: (i) decreasing DE
state parameters $w < -1$ are met; (ii) upper limits on $M_\nu$ are
gradually softened. This is shown in Figure \ref{mnu}.

The (i) point is consistent with the results of the previous Section.
The only difference being that the dataset considered here includes
SNIa, instead of $H_0$ data. 

As far as the (ii) point is considered, we show the effects of taking
two specific $\Omega_b h^2$ values. The former one is the value
corresponding to the fit at $w = -1.8$, discussed in the previous
Section. The slight discrepancy from the $w$ range found here arises
from two reasons: (i) $H_0$ data are replaced by SNIa data. (ii)
Different parameters are set free: here $\Omega_b h^2$ is fixed and
$f_\nu$ (the fraction of Dark Matter due to $\nu$'s) is free, the
opposite choice with respect to Figure \ref{datw1}.

The meaning of the Figure is however clear: If $^7 Li$ data are
approached, $\nu$-mass constraint soften. If we opt for the closer
model allowing some agreement with $^7 Li$, the upper limit on $M_\nu$
shifts from 1.12$\, $eV to 1.27$\, $eV. If we go as down as is needed
to meet the g.a.~line, we find a limit at 1.56$\, $eV. Furthermore, in
the latter case, there is an apparent 1.7~$\sigma$'s ``detection'' for
$M_\nu$. Such pseudo-detection would take a completely different
significance if particle data would yield a neutrino mass detection in
a similar mass range.

\section{Discussion}
It may be worth outlining soon that {\it phantom} DE is not a
straightforward result of quantum field theory. If DE is a scalar
field, its energy density $\rho_{DE}$ and pressure $p_{DE}$ arise from
the kinetic and potential energy densities ($E_k$ and $V$,
respectively) according to the relations
\begin{equation}
\rho_{DE} = E_k + V~,~~~~ p_{DE} = E_k - V~,
\end{equation}
so that $w = p_{DE}/\rho_{DE}$ approaches $-1$ when $E_k \ll V$, but is
unable to bypass the $w = -1$ limit unless $E_k$ lays in the interval
$-V < E_k < 0$.

\begin{figure}
\begin{center}
\includegraphics[height=9.cm,angle=0]{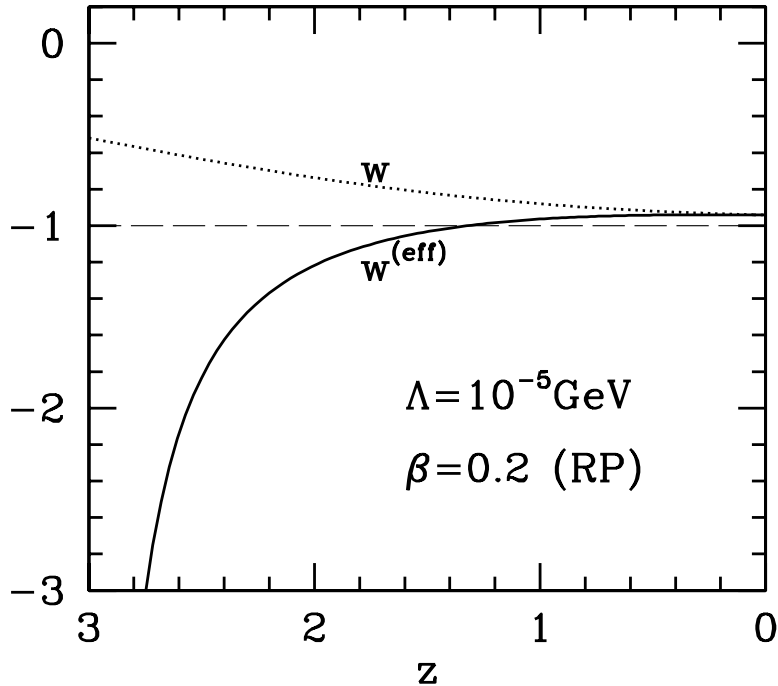}
\end{center}
\caption{Redshift dependence of the (effective) state parameter
in a dynamical DE model.}
\label{RP}
\end{figure}
Another option is that data analysis tends to favor $w < -1$ because
Cold Dark Matter (CDM) and DE are coupled and the assumed parameter
space does not include a coupling parameter. In other terms: measuring
$w \ll -1$ could be the consequence of assuming no energy flow from
CDM to DE if they are actually coupled.

In this case, data can be approached by using an effective DE energy
density \cite{das}
\begin{equation}
\rho^{(eff)}_{DE} = \rho_{DE} + \rho_{CDM} (1+z)^3 [f(\phi)-1]
~~~~~{\rm with} ~~~~ C = df/d\phi
\end{equation}
being the CDM-DE coupling (see, e.g., \cite{amendola00}). The state
parameter of DE reads then
\begin{equation}
w^{(eff)} = \frac{w}{1-x}
\end{equation}
with
\begin{equation}
x = (\rho_{CDM}/\rho_{DE}) [1 - e^{-C(\phi_0-\phi)}],
\end{equation}
provided that $C = (\beta/m_p) \times 4 \sqrt{\pi/3}$ is constant. It
must be however outlined that this is however an approximate
treatment, yielding increasingly rougher results as the coupling
increases. In Figure \ref{RP} we give an example of the redshift
dependence of $w^{(eff)}$, for a RP model \cite{RP88} with
cosmological parameters close to WMAP7 best fit ($\Omega_ch^2=0.114$,
$\Omega_bh^2 = 0.0226$, $h=0.7$, $\Lambda$ and $\beta$ shown in the
frame). In this straightforward case, we find a variable effective
state parameter which, averaged on $z$ between $z=0$ and $z=2.5$,
yields $\langle w^{(eff)} \rangle \sim -1.18~.$ As a matter of fact,
however, models yielding significantly more negative $\langle
w^{(eff)} \rangle$ averages are hard to build, unless assuming that
$\beta$ itself suitably depends on $z~.$ The point is that, when
increasing $\beta$, there arise apparent instabilities in $w^{(eff)}$,
even below $z=2~.$ On the contrary, coupled DE models with similar
$\beta$'s, independently of their capacity to meet data, certainly
exhibit no such instabilities. Henceforth, the capacity to mimic
CDM-DE coupling by means of an effective potential is limited.

However, if $w \ll -1$ is suggested by $^7 Li$ data, this can be
generically interpreted as possibly favoring CDM-DE coupling.

In recent work, the option of CDM-DE coupling has been specifically
tested against data, by simultaneously allowing for higher $M_\nu$
\cite{lavacca09,kristiansen09}, finding that such Mildly Mixed Coupled
(MMC) models exhibit a likelihood slightly exceeding $\Lambda$CDM
(below the 2-$\sigma$ level). We plan to reconsider such option in the
presence of priors towards lower $\Omega_b h^2$.

\section{Conclusions}
The main task of this note is testing the consequence of taking $^7
Li$ data on the same foot of other light nuclide abundances. BBN
constraints, when $^7 Li$ is disregarded, yield no real parameter
limitation in top of those arising from CMB, BAO, SNIa, $H_0$, etc.~:
the BBN interval set by $^4 He$, $^3 He$, and $^2 H$, sometimes set as
a prior, more or less overlaps with WMAP interval for $\Lambda$CDM
cosmologies.

On the contrary, if $^7 Li$ is considered, BBN is a real extra
constraint to cosmological models and DE state equations with $w \ll
-1$ are favored together with $\Omega_b h^2$ values smaller by $\sim
4$-$5\, \%$, in respect to the interval usually considered.

Within this context, $M_\nu$ limits are softened up to 1.6$\, $eV, but
the really new feature is a quasi-detection for a non-vanishing $M_\nu
\sim 0.9 \pm 0.5~$eV.

A more general, very tentative, conclusion is that there is a
realistic possibility that WMAP data, for some still unknown reason,
led to slightly overestimate $\Omega_b$; if such estimate is reduced
by 4-5$\, \%$, then, a different cosmological scenario seems to open:
a simple $\Lambda$CDM cosmology is no longer so close to data,
predictions of light element abundances by BBN would approach a
general self-consistency, an energy flow from CDM to DE could mimic a
unique nature of the dark components, and $M_\nu \sim 0.9~$eV could
turn into an actual detection.

\acknowledgments The authors wish to thank F. Matteucci for Figure
\ref{matteucci} and for insightful discussions. This work was
partially supported by ASI (the Italian Space Agency) thanks to the
contract I/016/07/0 "COFIS". S.B. acknowledges the support of the
Italian Center for Space Physics (CIFS) through the C.I. no.~2010/24~.

\end{document}